\documentclass[letterpaper,pra,twocolumn,showpacs,superscriptaddress]{revtex4}
\usepackage{graphicx}
\usepackage{amsmath}
\usepackage{amssymb}
\usepackage{dsfont}

\newcommand{\ket}[1]{|#1\rangle}
\newcommand{\bra}[1]{\langle #1|}

\begin{document}
\title{Quantum Cloning of Binary Coherent States \\ --  Optimal Transformations and Practical Limits --}

\author{Christian~R.~M\"{u}ller}	
\affiliation{Max Planck Institute for the Science of Light, G\"unther-Scharowsky Str. 1 Bldg. 24, D-91058 Erlangen, Germany}
\affiliation{Institute of Optics, Information and Photonics, University of Erlangen-Nuremberg, Staudtstr. 7/B2, D-91058 Erlangen, Germany}
\affiliation{Department of Physics, Technical University of Denmark, Fysikvej, 2800 Kgs. Lyngby, Denmark}

\author{Gerd~Leuchs}							
\affiliation{Max Planck Institute for the Science of Light, G\"unther-Scharowsky Str. 1 Bldg. 24, D-91058 Erlangen, Germany}
\affiliation{Institute of Optics, Information and Photonics, University of Erlangen-Nuremberg, Staudtstr. 7/B2, D-91058 Erlangen, Germany}
\affiliation{Department of Physics and Max Planck - University of Ottawa Centre for Extreme and Quantum Photonics, \\ University of Ottawa , 25 Templeton, Ottawa ON K1N 6N5, Canada}

\author{Christoph~Marquardt}							
\affiliation{Max Planck Institute for the Science of Light, G\"unther-Scharowsky Str. 1 Bldg. 24, D-91058 Erlangen, Germany}
\affiliation{Institute of Optics, Information and Photonics, University of Erlangen-Nuremberg, Staudtstr. 7/B2, D-91058 Erlangen, Germany}

\author{Ulrik~L.~Andersen}					  
\affiliation{Department of Physics, Technical University of Denmark, Fysikvej, 2800 Kgs. Lyngby, Denmark}
\affiliation{Max Planck Institute for the Science of Light, G\"unther-Scharowsky Str. 1 Bldg. 24, D-91058 Erlangen, Germany}
\affiliation{Institute of Optics, Information and Photonics, University of Erlangen-Nuremberg, Staudtstr. 7/B2, D-91058 Erlangen, Germany}

\email{christian.mueller@mpl.mpg.de}

\begin{abstract}
The notions of qubits and coherent states correspond to different physical systems and are described by specific formalisms. 
Qubits are associated with a two-dimensional Hilbert space and can be illustrated on the Bloch sphere. 
In contrast, the underlying Hilbert space of coherent states is infinite-dimensional and the states are typically represented in phase space. 
For the particular case of binary coherent state alphabets these otherwise distinct formalisms can equally be applied. 
We capitalize this formal connection to analyse the properties of optimally cloned binary coherent states. 
Several practical and near-optimal cloning schemes are discussed and the associated fidelities are compared
to the performance of the optimal cloner.
\end{abstract}
\maketitle

\section*{Introduction}
Classical communication and data processing are based on the transmission and manipulation of two distinct
states usually termed 'ON' and 'OFF' or '1' and '0'. 
In quantum physics, the formalism is enriched by the superposition principle and the set of possible signals is extended to the realm of non-orthogonal states. 
Qubits \cite{Kok2007} are quantum mechanical two level systems such that any set of qubits can be described in a two-dimensional Hilbert space spanned by two basis vectors $\ket{0}$ and $\ket{1}$.
Coherent states are the quantum states whose dynamics most closely resemble the undulatory characteristic of a classical plain wave, i.e. of a classical harmonic oscillator. 
They are readily produced, loss tolerant and have proven to be ideal signals carriers in conventional telecommunication. Moreover, coherent states are of outstanding importance in novel applications such as quantum repeaters \cite{VanLoock2006, VanLoock2008} quantum communication \cite{Heim2014} or quantum key distribution (QKD) \cite{Grosshans2002,Inoue2003}. \newline
A coherent state $\ket{\alpha}$ \cite{Braunstein2005, Weedbrook2012, Adesso2014} is a pure state ans can thus be described by a single Hilbert space vector. 
In contrast to qubit states, however, the underlying Hilbert space of the coherent states is infinite dimensional. 
The description of a set of $N$ coherent states generally requires a Hilbert space of dimension $N$ or larger.
For binary coherent states (BCS) $\{ \ket{\pm\alpha} \}$ we have $N=2$ such that this alphabet can be described in a two-dimensional Hilbert space.
Therefore, while typically represented within the formal framework of continuous variables, the BCS alphabet can equally be described in a qubit-like fashion. 

In this Article we put the formal analogy between binary coherent states and qubit states into practice to analyze the performance of the optimal quantum cloning strategy for the BCS alphabet.
Several practical and near-optimal quantum cloning strategies are then described and their performance is compared to the optimal cloner. 
In Sec.\ref{Bases} we illustrate the formal connection between the description of binary coherent states and qubit states by representing the BCS in different two-dimensional bases. 
In Sec. \ref{Discrimination} we review different quantum receivers for the discrimination of binary coherent states. 
These receivers are later considered as components of more complex quantum cloning schemes. 
Sec. \ref{Cloning} is concerned with the cloning of binary coherent states. 
We investigate the performance of different strategies and compare the achievable fidelity with the bound obtained for an
optimal state dependent qubit cloner. \newline

\section{Two Dimensional Representations of the Binary Coherent States}\label{Bases}
Qubits can be described as superpositions of two orthonormal basis states spanning a two-dimensional Hilbert space. 
For coherent states $\ket{\alpha}$, the underlying Hilbert space is in general infinite-dimensional and the states demand a distinctly different description. 
This is exemplified by the representation of coherent states in terms of an infinite superposition of photon number states.
\begin{equation}
\ket{\alpha} =  \exp\left(-\frac{|\alpha|^2}{2}\right) \sum_{n}{\frac{\alpha^{n}}{\sqrt{n!}} \ket{n}}
\label{eq:coherent_state_formal}
\end{equation}
As we are interested in binary coherent states, we can confine the Hilbert space to a two dimensional subspace spanned, for instance, by the two state vectors $\ket{\pm \alpha}$.
Therefore, we can choose a suitable basis for the task at hand and represent the BCS with methods borrowed from the qubit formalism.
In the following we will discuss two important bases.
The first basis is a direct generalization of the conventional representation of qubits in terms of two non-orthogonal basis states and we will refer to it as the \textit{qubit basis}. 
The second basis represents the BCS in terms of Schr\"{o}dinger cat/kitten states and we refer to this basis as the \textit{cat-state basis} \cite{Knight1997, Jeong2002, Jeong2007}.

\begin{figure}[tb]%
\includegraphics[width=1\columnwidth]{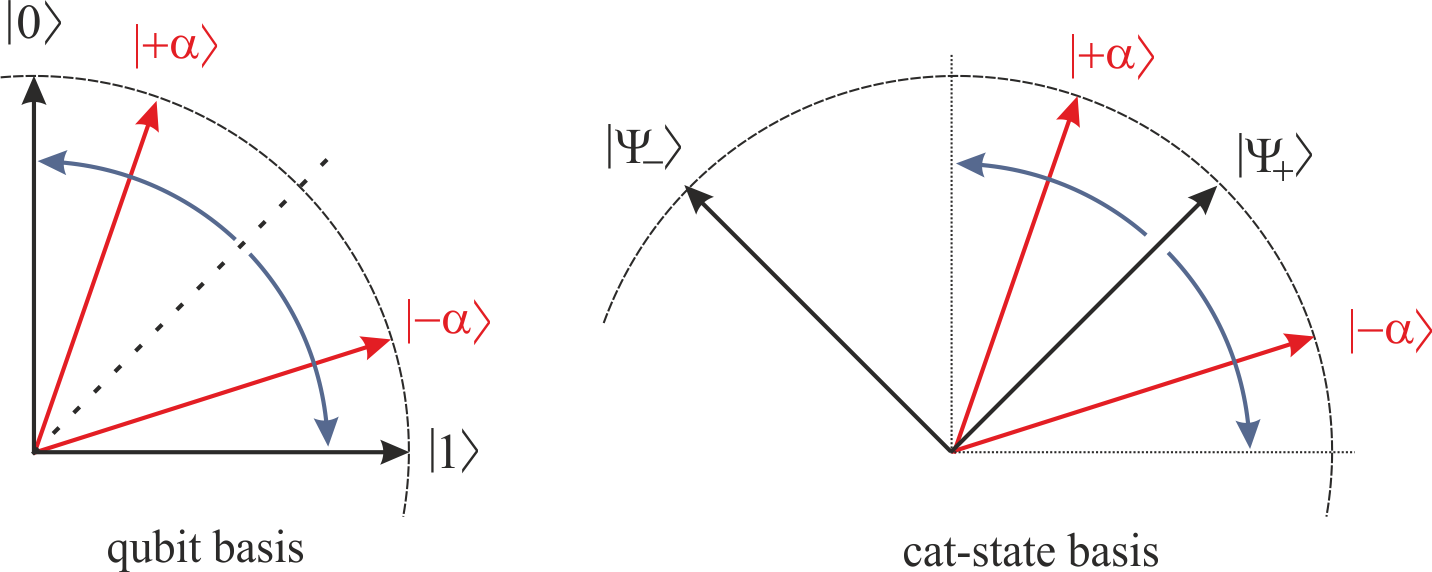}%
\caption{Hilbert space representation of the coherent signal states and the different basis states. 
(left) qubit basis: the signal states approach the $45^{\circ}$ diagonal in the limit of vanishing coherent amplitude $|\alpha|\mapsto 0$, and coincide with either of the basis states in the classical limit $|\alpha|\gg1$. 
(right) cat-state basis:  the situation is reversed. 
The signal states encompass an angle of $45^{\circ}$ with respect to the basis states in the classical limit and coincide with $\ket{\Psi_{+}}$ for $|\alpha|\mapsto 0$.}
\label{HilbertSpaceVectors}%
\end{figure}

\subsection*{Qubit Basis} 
A suitable parametrization of BCS in terms of two orthonormal basis states $\ket{0}$ and $\ket{1}$ is founded in an \textit{overlap angle} $\theta$, where $\sin{(2\theta)}=\bra{\alpha}-\alpha\rangle$
\begin{eqnarray}
\ket{+\alpha} &=& \cos(\theta) \ket{0} + \sin(\theta) \ket{1} \\
\ket{-\alpha} &=& \sin(\theta) \ket{0} + \cos(\theta) \ket{1}, \,\,\, \theta \in \left[ 0, \frac{\pi}{4}\right].  \nonumber
\label{eq:qubit_representation}
\end{eqnarray}
The corresponding basis states in terms of the binary coherent states are given by 
\begin{eqnarray}
\ket{0} &=& \frac{\cos(\theta)\ket{+\alpha} - \sin(\theta)\ket{-\alpha}}{\cos(2\theta)} \\
\ket{1} &=& \frac{-\sin(\theta)\ket{+\alpha} + \cos(\theta)\ket{-\alpha}}{\cos(2\theta)},   \nonumber
\label{eq:qubit_representation2}
\end{eqnarray}
where the factor in the denominator accounts for the non-orthogonality of the coherent states and ensures normalization.

Qubits can also be represented on the Bloch sphere.
In the general case, i.e. including mixed states, the qubit density matrix $\rho$ can be expressed in terms of the Pauli spin matrices and the unit operator
\begin{equation}
\rho = \frac{1}{2}(\mathds{1}+x\,\textbf{$\sigma$}_x+y\,\textbf{$\sigma$}_y+z\,\textbf{$\sigma$}_z)= \frac{1}{2}\left(
\begin{matrix}
1+z&x-iy\\
x+iy&1-z
\end{matrix}\right), 
\label{eq:qubit_density_matrix}
\end{equation}
and the amplitudes $x$, $y$, and $z$ are the Cartesian coordinates describing the position in or on the Bloch sphere. 

In the $\theta$-parametrized qubit basis, the BCS density operator is expressed as
\begin{eqnarray}
\rho &=& \frac{1}{2}(\ket{+\alpha}\bra{+\alpha}+\ket{-\alpha}\bra{-\alpha}) \\
					 &=& \frac{1}{2}(|0\rangle\langle0|+|1\rangle\langle1|+\sin(2\theta)|0\rangle\langle1|+\sin(2\theta)|1\rangle\langle 0|) \nonumber \\
					 &=& \frac{1}{2}
\left(\begin{matrix}
1&\sin(2\theta)\\
\sin(2\theta)&1
\end{matrix}\right), \nonumber
\label{QubitDensityOperator}
\end{eqnarray}
and the coordinates of the Bloch vector \textbf{s} are  

\begin{align}
x = \sin(2\theta) ,\hspace{0,5cm}
y = 0 ,\hspace{0,5cm}
z = 0.
\label{QubitPauliComponentsBoth}
\end{align}

The density matrices and Bloch sphere coordinates of the individual signal states follow as 
\begin{eqnarray}
\ket{+\alpha}\bra{+\alpha} = \frac{1}{2}\left(\begin{matrix}
2\cos^2(\theta)&\sin(2\theta)\\
\sin(2\theta)&2\sin^2(\theta) 
\end{matrix}\right) \\
\ket{-\alpha}\bra{-\alpha} = \frac{1}{2}\left(\begin{matrix}
2\sin^2(\theta)&\sin(2\theta) \nonumber\\
\sin(2\theta)&2\cos^2(\theta) 
\end{matrix}\right) \\
\Rightarrow x = \sin(2\theta) ,\hspace{0,5cm} y = 0 ,\hspace{0,5cm} z = \pm\cos(2\theta) \nonumber
\label{QubitPauliComponents}
\end{eqnarray}
A Hilbert space representation of the qubit basis vectors and the binary coherent states are depicted in Fig.\ref{HilbertSpaceVectors}a). 

\subsection*{Cat-State Basis} 
An alternative representation of the BCS is offered by the in-phase and out-of-phase superposition of the coherent signal states $\Psi_{\pm}$, which are mutually orthogonal and commonly referred to as even and odd (Schr\"{o}dinger) cat-states. 
Including the appropriate normalization constants, $\Omega_{\pm} = \sqrt{1\pm e^{-2|\alpha|^2}}$, the cat-state basis states are

\begin{eqnarray}
\ket{\Psi_{+}} = \frac{1}{\sqrt{2} \Omega_{+}} \left(\ket{+\alpha} + \ket{-\alpha} \right), \\
\ket{\Psi_{-}} = \frac{1}{\sqrt{2} \Omega_{-}} \left(\ket{+\alpha} - \ket{-\alpha} \right).\nonumber
\label{cat state}
\end{eqnarray}

The coherent states are expressed in the cat-state basis as
\begin{eqnarray}
\ket{+\alpha} = \frac{1}{\sqrt{2}} \left( \Omega_{+} \ket{\Psi_{+}}  +  \Omega_{-} \ket{\Psi_{-}} \right), \\
\ket{-\alpha} = \frac{1}{\sqrt{2}} \left( \Omega_{+} \ket{\Psi_{+}}  -  \Omega_{-} \ket{\Psi_{-}} \right). \nonumber
\label{CohInCatBas}
\end{eqnarray}

The density matrix of the BCS with equal prior probabilities $\rho = \frac{1}{2}\left( \ket{+\alpha}\bra{+\alpha} + \ket{-\alpha}\bra{-\alpha}  \right)$ is diagonal in the cat-state basis
\begin{eqnarray}
\rho &=& \frac{1}{2} \left({\begin{array}{cc} \Omega_{+}^{2} & 0 \\  0 & \Omega_{-}^{2} \end{array}} \right) \\
     &=& \frac{1}{2} \left({\begin{array}{cc} 1+e^{-2|\alpha|^2} & 0 \\  0 & 1-e^{-2|\alpha|^2} \end{array}} \right), \nonumber
\label{RhoFullState}
\end{eqnarray}
while the density matrices of the individual states take the form 

\begin{eqnarray}
\rho_{\pm\alpha} &=& \frac{1}{2} \left({\begin{array}{cc} \Omega_{+}^{2} & \pm\Omega_{+}\Omega_{-}\\ \pm\Omega_{-}\Omega_{+}& \Omega_{-}^{2} \end{array}} \right) \\
								 &=& \frac{1}{2} \left({\begin{array}{cc} 1+e^{-2|\alpha|^2} & \pm\sqrt{1-e^{-4|\alpha|^2}} \\ \pm \sqrt{1-e^{-4|\alpha|^2}} & 1-e^{-2|\alpha|^2} \end{array}} \right). \nonumber
\label{RhoIndividualState}
\end{eqnarray}

The Hilbert space representation of the cat-state basis vectors and the signal states are depicted in Fig.\ref{HilbertSpaceVectors}b). 
This basis is rotated by $45^{\circ}$ with respect to the qubit basis and the signal states are symmetrically aligned with respect to the basis state $\ket{\Psi_{+}}$. In the classical limit $|\alpha|\gg1$, the signal states are aligned on the $\pm 45^{\circ}$ diagonals with respect to the basis vectors. 

\begin{figure}%
\includegraphics[width=1\columnwidth]{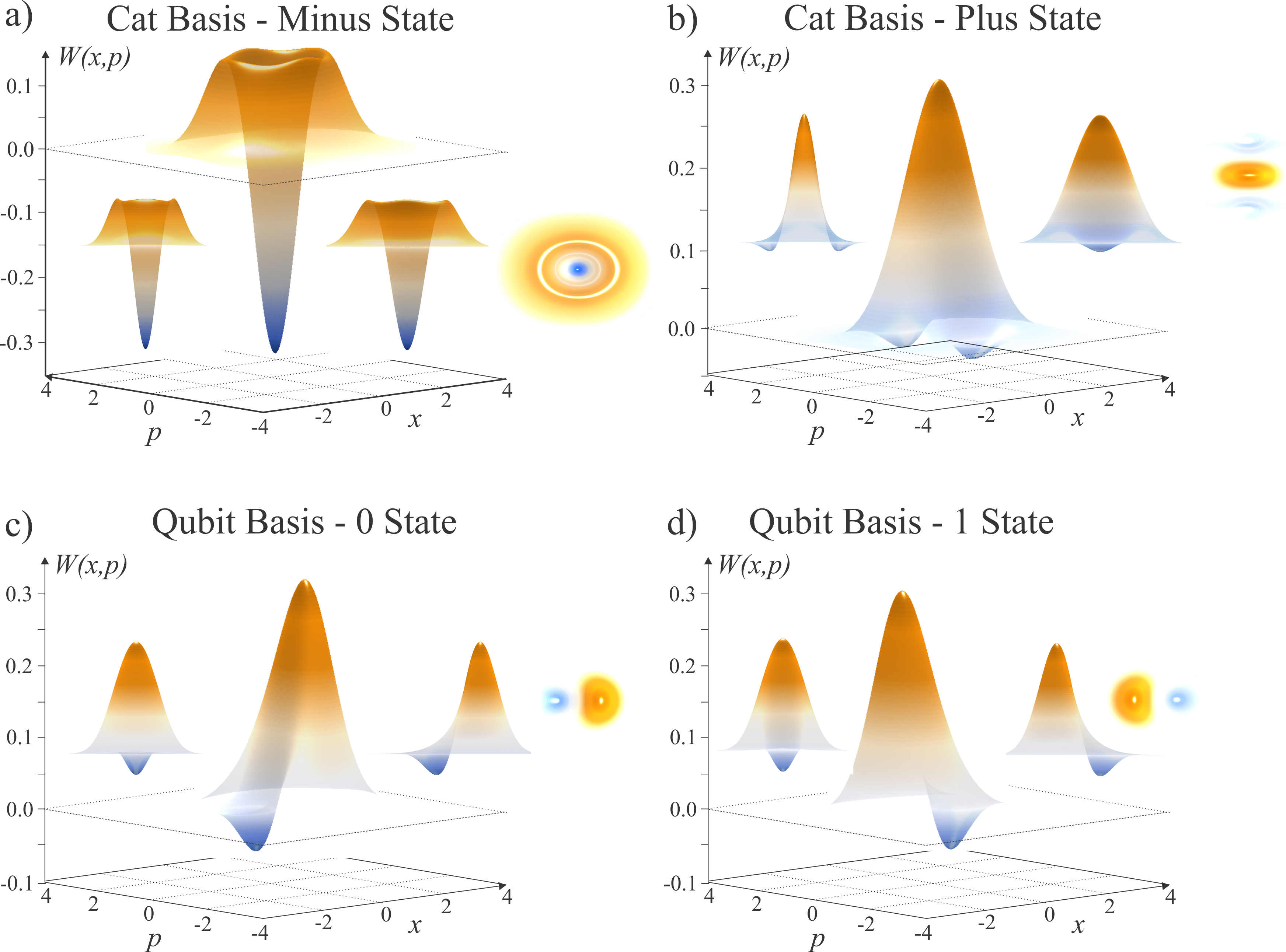}%
\caption{ 
Illustrations of the BCS basis states in terms of their Wigner functions. 
A perspective view is shown by the big illustration in the coordinate frame. 
The smaller illustrations depict, from left to right, a view along the \textit{p}-quadrature, a view along the \textit{x}-quadrature, and a view from the top. 
a)-b) Wigner function representations of the odd and even cat-basis states.
c)-d) Representations of the qubit basis states. The Wigner functions exhibit mirror symmetry. }%
\label{BasisStatesCollection}%
\end{figure}

An illustrative alternative to the Hilbert space representations is to depict the Wigner functions of the different basis states. 
These are shown from different perspectives in Fig.\ref{BasisStatesCollection}. 

\section{Discrimination of Binary Coherent States} \label{Discrimination}
A \textit{classical} cloner or copying machine can be described as a device that performs an adequate measurement on the object to be copied and subsequently prepares a replica based on the information retrieved during the measurement process.
For the BCS this measurement merely needs to discriminate between the two input states.
However, it is one of the innermost consequences of the laws of quantum mechanics that non-orthogonal states can not be discriminated with certainty.
Optimal detection strategies were first investigated by Helstrom \cite{Helstrom1967, Helstrom1976} and Holevo \cite{Holevo1982} and a lot of attention has since been devoted to the development of optimal and near-optimal receivers for binary coherent states \cite{Dolinar1973, Kennedy1973, Hirota96, HirotaBan96, Takeoka2008, Wittmann2008, Wittmann2010, Wittmann2010b, Nair12, Banaszek99} and for the discrimination of larger signal alphabets \cite{Assalini11, Becerra11, Mueller12, Izumi12, Becerra13NatPhot, Becerra13NatComm, Nair14, Becerra14, Mueller15}.

In principle, one can differentiate between two fundamental discrimination strategies:
minimum error discrimination (MED) and unambiguous state discrimination (USD) \cite{Ivanovic1987, Dieks1988, Peres1988}.
In MED the receiver is tailored to minimize the average error probability. The strategy is deterministic such that a meaningful hypothesis is assigned to each individual signal state.
USD, in contrast, is a probabilistic state discrimination strategy. 
It allows to perfectly discriminate also non-orthogonal quantum states at the expense of a finite probability for an inconclusive result $p_{inc}$ that does not provide any information about the state \cite{Huttner1996}. 
An intermediate regime where both erroneous and inconclusive results are allowed has also been considered \cite{Wittmann2010} and the minimal probability of error for a fixed probability of inconclusive results has been derived for pure \cite{Chefles1998} and mixed states \cite{Fiurasek2003}.
In the following, we always assume equal prior probabilities for the signal states which is also the typical case in classical communication. 

\begin{figure}%
\includegraphics[width=.9\columnwidth]{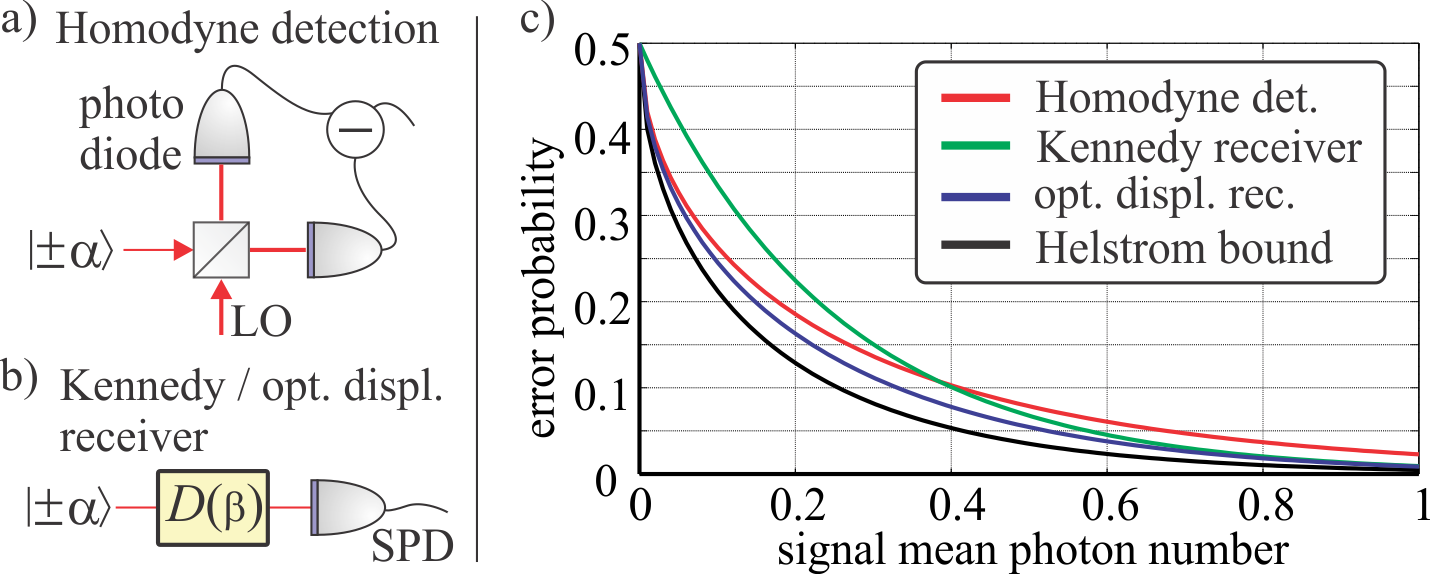}%
\caption{
a) Sketch of a homodyne detector. The signal state $\ket{\pm\alpha}$ interferes with a bright local oscillator (LO) at a symmetric beam splitter and the emerging beams are detected by pin photo diodes. 
The difference signal yields a projected quadrature value where the quadrature phase is determined by the phase of the LO. 
b) Sketch of the Kennedy receiver $(|\beta|=|\alpha|)$ and the optimized displacement receiver ($|\beta|$ optimized).
The signal state is coherently displaced in phase space and is subsequently detected by a single photon detector (SPD). 
c) Comparison of the error probability between the different receivers \cite{Kennedy1973,Wittmann2010} and the Helstrom bound.}%
\label{error_probability}%
\end{figure}
In order to obtain the correct hypothesis with high probability, an adequate measurement needs to be performed. 
The standard quantum limit for the discrimination of binary coherent states is defined as the minimal error probability that can be obtained via direct measurement of the encoding variable.
For the binary coherent state this is the in-phase-quadrature component which can be measured via homodyne detection, see Fig.\ref{error_probability}a). 
The hypothesis $H_{\pm}$ associated with the state $\ket{\pm\alpha}$ is determined by the sign of the measured quadrature value. 
\newpage
Positive values map to $\ket{\alpha}$, while negative values yield~$\ket{-\alpha}$.
\begin{eqnarray}
&H_{+}:& x\geq0 \mapsto \ket{+\alpha}, \nonumber \\
&H_{-}:& x<0 \mapsto \ket{-\alpha}
\label{eq:HypothesisHD}
\end{eqnarray}
This strategy is described by two positive operator-valued measure (POVM) elements corresponding to projections onto the negative and positive quadrature semi-axis.
\begin{eqnarray}
\hat{\Pi}_{+}^{HD} &=& \int_{0}^{\infty}  \ket{x}\bra{x} \,\mathrm{d}x, \nonumber \\
\hat{\Pi}_{-}^{HD} &=& \int_{-\infty}^{0} \ket{x}\bra{x} \,\mathrm{d}x = \mathds{1} - \hat{\Pi}_{1}^{HD}.
\label{eq:HomodynePOVM}
\end{eqnarray}
The error probability of the homodyne approach is determined by the overlap of the signals' marginal distributions with the opposed quadrature semi-axis and amounts to
\begin{equation}
p_{err}^{HD}(\alpha) = \frac{1}{2} \left(1 - \mathrm{erf}(\sqrt{2}|\alpha|) \right).
\label{eq:ErrorProbabilityHD}
\end{equation}

From a purely classical perspective, this strategy is optimal. 
Yet, quantum mechanics allows for an even smaller error probability. 
This limit is called Helstrom bound \cite{Helstrom1967, Helstrom1976} and is determined by the overlap of the coherent states $|\bra{-\alpha}\,\alpha\rangle|^2 = e^{-4|\alpha|^2}$.
\begin{equation}
p_{err}^{H}(\alpha) = \frac{1}{2}\left(1 - \sqrt{1  - e^{-4|\alpha|^2}} \right)
\label{eq:HelstromBound}
\end{equation}

A receiver reaching the Helstrom bound and based on photon detection and instant feedback has been proposed by Dolinar \cite{Dolinar1973}.
A near-optimal receiver based on a fixed phase-space displacement followed by single photon detection has been proposed by Kennedy \cite{Kennedy1973} and was further developed to the optimized displacement receiver by Takeoka and Sasaki \cite{Takeoka2008, Wittmann2008}, see Fig.\ref{error_probability}b).
The displacement transforms the signals to a dim state close to the vacuum and to a coherent state of at least twice the original amplitude, respectively.
\begin{eqnarray}
\ket{-\alpha}&\mapsto& \hat{D}(\beta)\ket{-\alpha} = \ket{-\alpha+\beta},  \\
\ket{+\alpha}&\mapsto& \hat{D}(\beta)\ket{+\alpha} = \ket{+\alpha+\beta}, \nonumber
\label{eq:KennedyDisplacedStates}
\end{eqnarray}
where $\hat{D}(\beta)=\exp(\beta\hat{a}^{\dagger}-\beta^*\hat{a})$ is the displacement operator and $|\beta| = |\alpha|$ for the Kennedy receiver. 

Whenever the single photon detector does not register a photon, the hypothesis is $H_{-}$. 
Correspondingly, the hypothesis is $H_{+}$ if at least one photon was observed. 
The POVM elements incorporating the initial displacement and the subsequent photon number measurement correspond to projections onto the coherent state $\ket{-\beta}$ and its Hilbert space complement $\mathds{1} - \ket{-\beta}\bra{-\beta}$.
\begin{eqnarray}
\hat{\Pi}^K_{-} &=& \hat{D}^{\dagger}(\beta)\ket{0}\bra{0}\hat{D}(\beta) \\
&=& \ket{-\beta}\bra{-\beta}, \nonumber \\
\hat{\Pi}^K_{+} &=& \hat{D}^{\dagger}(\beta)\left( \sum_{n=1}^{\infty} \ket{n}\bra{n} \right) \hat{D}(\beta) \nonumber \\
&=& \hat{D}^{\dagger}(\beta) \left(\mathds{1} - \ket{0}\bra{0} \right) \hat{D}(\beta)  \nonumber \\
&=& \mathds{1} - \ket{-\beta}\bra{-\beta}. \nonumber 
\label{eq:KennedyPOVM}
\end{eqnarray}

In the Kennedy receiver the state displaced to the vacuum is always identified correctly, as the vacuum is an eigenstate of the photon number basis. 
Erroneous hypotheses are solely due to measurements in which the bright state failed to excite a photon detection such that the error probability amounts to $p_{err}^{K} =  \frac{1}{2} e^{-4|\alpha|^2}$.
In contrast, the optimized displacement receiver minimizes the total error probability optimizing over the displacement amplitude $\beta$, where the optimal parameter is given by the solution of the transcendental equation $\alpha = \beta\tanh(2\alpha\beta)$ and takes values $\beta\geq\frac{1}{\sqrt{2}}$.
The resulting error probability is 
\begin{equation}
p_{err}^{OD} =  \frac{1}{2} - e^{-(|\alpha|^2+|\beta|^2)}\sinh(2\alpha\beta).
\label{eq:optDisplErrorProb}
\end{equation}
The receiver can be further improved by squeezing the signal states \cite{Takeoka2008}. 
The error probabilities of the different receivers are shown as a function of the signal mean photon number in Fig.\ref{error_probability}c).

\section{Binary Coherent State Cloning}  \label{Cloning}
The no-cloning theorem epitomizes the basic tenets of quantum theory. 
It states that the preparation of perfect copies of an arbitrary unknown quantum state is impossible.
The theorem was originally formulated in terms of qubit systems \cite{Wooters1982, Dieks1982} but has been generalized to the regimes of continuous variables as well \cite{Cerf2000, Cerf2000a}.
The objective of quantum cloning \cite{RevModPhys.77.1225} is to maximize the overlap of the clones $\{\hat{\rho}_{k}\}$ to the input state $\{\ket{\psi_{k}}\}$.
The performance can be quantified by the mean fidelity F.
\begin{equation}
\mathrm{F} = \sum_{k} p_{k} \bra{\psi_{k}} \hat{\rho}_{k} \ket{\psi_{k}}, \,\,\, \sum_{k} p_{k} = 1, 
\label{eq:fidelity}
\end{equation}
where $p_{k}$ are the \textit{a priori} probabilities of the individual states. 
In the following, we restrict the analysis to BCS with \textit{equal priors} $p_{\pm} = \frac{1}{2}$. 
Quantum information theory readily offers the tools to maximize the fidelity in terms of optimized unitary operations. 
Translating the theoretical results to an experimental implementation, however, is in general a non-trivial task.
In the following sections, we first derive the density matrix and characteristic properties of optimally cloned binary coherent states. 
Subsequently, we describe different near-optimal cloning schemes and compare their performance to the optimal BCS cloner.

\subsection{Optimally Cloned Binary Coherent States}
The optimal fidelity for state-independent cloning was shown to be  $\mathrm{F}=\frac{5}{6}$ for qubits \cite{Buzek1996} and $\mathrm{F}=\frac{2}{3}$ for coherent states \cite{Cerf2000a}. For the particular case of binary qubit states, an upper bound for the cloning fidelity has been derived by~Bru\ss\,\textit{et~al.}~\cite{Bruss1998}.
\begin{equation}
\mathrm{F} \leq \frac{1}{2} \left( 1 + \frac{1-S^2}{\sqrt{1+S^2}} + \frac{S^2(1+S)}{1+S^2}\right).
\label{eq:maxFidBruss}
\end{equation}
The parameter $S = \sin{(2\theta)}$ 
denotes the initial overlap of the signal states in terms of the $\theta$-parametrized description of the qubit states, see Eq.(\ref{eq:qubit_representation}).  
Bru\ss\,\textit{et al.} also provide the transformations of the Bloch sphere coordinates under an optimal BCS cloning procedure. 
The transformation can be decomposed into two steps:
\begin{itemize}
	\item The modulus of the Bloch vector $|\textbf{s}|$ is reduced. Thus, the clones are in a mixed state.
		\begin{equation}
		|\textbf{s}| = \sqrt{\frac{S^2(1+S^2)}{(1+S^2)^2} + \frac{1-S^2}{1+S^2}}
		\label{eq:BlochLenghtRescale}
		\end{equation}
	\item The angle between the states in the Bloch sphere is reduced by an angle $\zeta$ and hence the mutual overlap of the clones is increased.
	\begin{equation}
	\zeta = \arccos\left( \frac{1}{|\textbf{s}|} \frac{\sqrt{1-S^2}}{\sqrt{1+S^2}}\right) - 2\theta
	\label{eq:BlochAngleChange}
	\end{equation}
\end{itemize}

\begin{figure}%
\includegraphics[width=1\columnwidth]{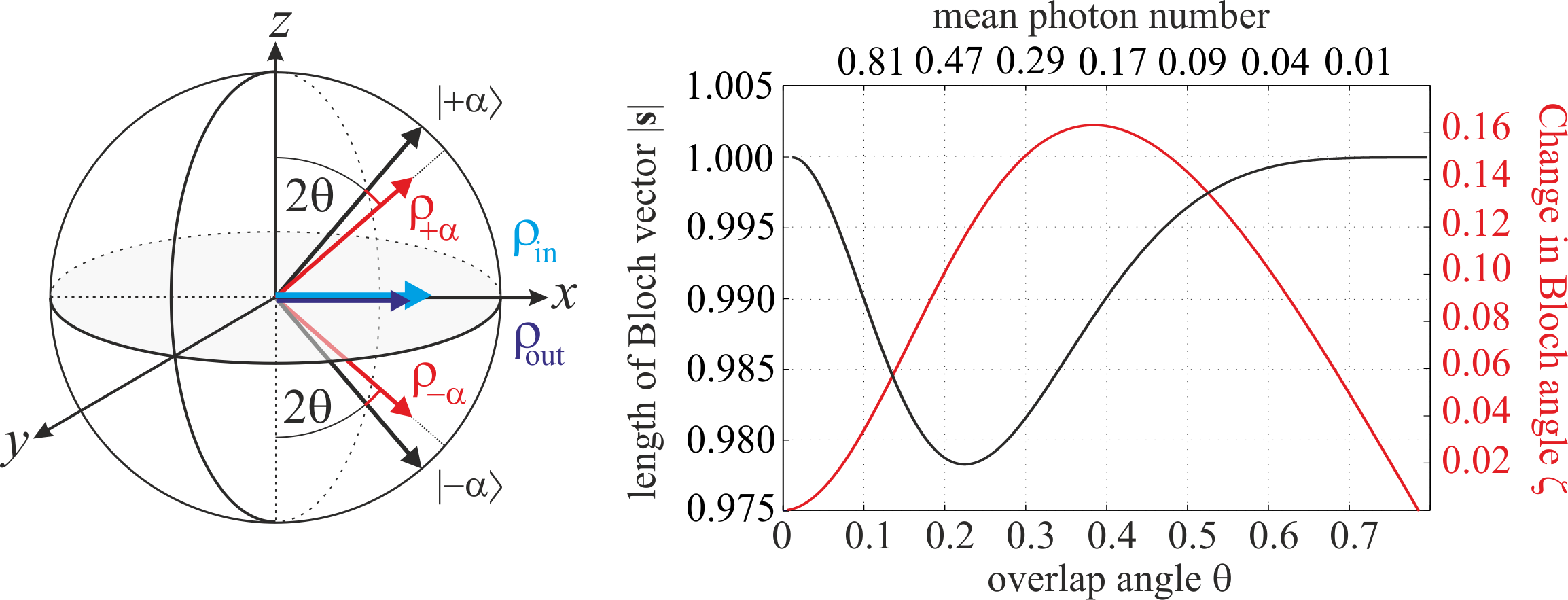}%
\caption{left) Bloch sphere representation of the input BCS and the resulting clones. 
right) The state transformations of the optimal binary state cloning procedure can be decomposed into two parts. 
First, the lengths of the Bloch vector are reduced, i.e. the clones are in a mixed state. 
Second, the angle between the vectors is reduced, i.e. the mutual overlap of the clones increases.}%
\label{BlochSphere}%
\end{figure}
This translates to a change in the probabilities of the Pauli matrix components, see Eq.(\ref{eq:qubit_density_matrix}),
\begin{eqnarray}
x' &=& \cos(2\theta + \zeta) \,|\textbf{s}|, \\
y' &=& y = 0, \nonumber \\
z' &=& \sin(2\theta + \zeta)\, |\textbf{s}|.\nonumber
\label{eq:BlochCoordinatesCloned}
\end{eqnarray}
The reduction of the Bloch vector norm and of the relative angle on the Bloch sphere are illustrated in Fig.\ref{BlochSphere} as a function of the  input state amplitude. 
In absolute quantities, these changes are small. 
The maximal reduction of the Bloch vector reduces its norm to about $|s|=0.978$ at an overlap angle of $\theta\approx0.225$, corresponding to a mean photon number $n=0.41$ for the individual signal states. 
The maximal change in $\theta$ is about $0.163$ rad at the overlap angle $\theta\approx0.38$ corresponding to $n=0.186$.

In order to derive the density matrix of the optimally cloned BCS we attempt the following strategy.
First, the BCS are described in the two-dimensional qubit-basis, see Eq.(\ref{eq:qubit_representation}).
Subsequently, the two-dimensional density matrix, see Eq.(\ref{QubitDensityOperator}), and the associated Bloch sphere representation, see Eq.(\ref{eq:qubit_density_matrix}), is deduced. 
The transformations invoked by the optimal binary qubit state cloning procedure, see Eq.(\ref{eq:BlochCoordinatesCloned}), are applied, and finally, the transformed density matrix is mapped back to the infinite dimensional Hilbert space of the coherent state basis in which we study the characteristics of the clones. 

In the extremal cases $\theta = 0$ (orthogonal states, $|\alpha|^2\gg1$) and $\theta = \frac{\pi}{4}$ (identical states, $|\alpha|^2=0$), the signals can be cloned perfectly. 
In any other case, the cloning is inevitably defective.
The minimal fidelity is $\mathrm{F}\approx98.54\%$ and is obtained at $\theta \approx 0.267$\,rad. 
The maximal fidelity as a function of the signal mean photon number is shown in Fig.\ref{CloningFidelity}.
At this point, the overlap of the coherent state wave functions is $|\langle -\alpha\mid\alpha\rangle|^2\approx0.259$ and the amplitude of the binary coherent state is $|\alpha| \approx0.581$.

In the following we compare the properties of the optimally cloned BCS to those of the coherent input signals. 
With respect to the symmetry of the alphabet, it suffices to restrict the analysis to either of the signal states $\ket{\alpha}$.

First, we translate the density matrix from the orthogonal qubit basis into the non-orthogonal basis of the binary coherent states.
\begin{align}
\rho' = \frac{1}{2} \left(\begin{matrix}
1+z'  & x'+iy'    \\
x'-iy'& 1-z' 
\end{matrix}\right)_{(|0\rangle,|1\rangle )} = \left(\begin{matrix}
\rho_{++} &\rho_{-+}   \\
\rho_{+-}& \rho_{--}
\end{matrix}\right)_{(|\pm \alpha \rangle )} 
\end{align}
where the coefficients are 
\begin{eqnarray}
\rho_{++} &=& \frac{z \cos(2\theta)+1-x\sin(2\theta)}{2\cos^2(2\theta)}, \nonumber\\
\rho_{+-} &=& \frac{x - \sin(2\theta)}{2\cos^2(2\theta)}, \nonumber\\
\rho_{-+} &=& \frac{x - \sin(2\theta)}{2\cos^2(2\theta)} = \rho_{+-}, \nonumber \\
\rho_{--} &=& \frac{z \cos(2\theta)-1+x\sin(2\theta)}{2\cos^2(2\theta)} .
\end{eqnarray}
Note, that this density matrix does in general not fulfill the condition for unit trace $\mathrm{Tr}[\rho]=1$, as the coherent states do not form a complete but an over-complete basis. 

Inserting e.g. the Bloch sphere coordinates of the coherent input state $\ket{+\alpha}$, see Eq.(\ref{QubitPauliComponents}), yields $\rho_{++}=1$ with all other matrix elements being zero. 
After the optimal cloning procedure, the Bloch sphere coordinates are modified according to Eq.(\ref{eq:BlochCoordinatesCloned}) such that
\begin{eqnarray}
\rho_{++} &=& \frac{|s|\cos(4\theta+\zeta)+1}{2\cos^2(2\theta)}, \\
\rho_{+-} &=& \frac{|s|\sin(2\theta+\zeta) - \sin(2\theta)}{2\cos^2(2\theta)}, \nonumber\\
\rho_{-+} &=& \frac{|s|\sin(2\theta+\zeta) - \sin(2\theta)}{2\cos^2(2\theta)} = \rho_{+-}, \nonumber \\
\rho_{--} &=& \frac{|s|\cos(\zeta) -1}{2\cos^2(2\theta)} \nonumber.
\end{eqnarray}

\begin{figure}%
\includegraphics[width=\columnwidth]{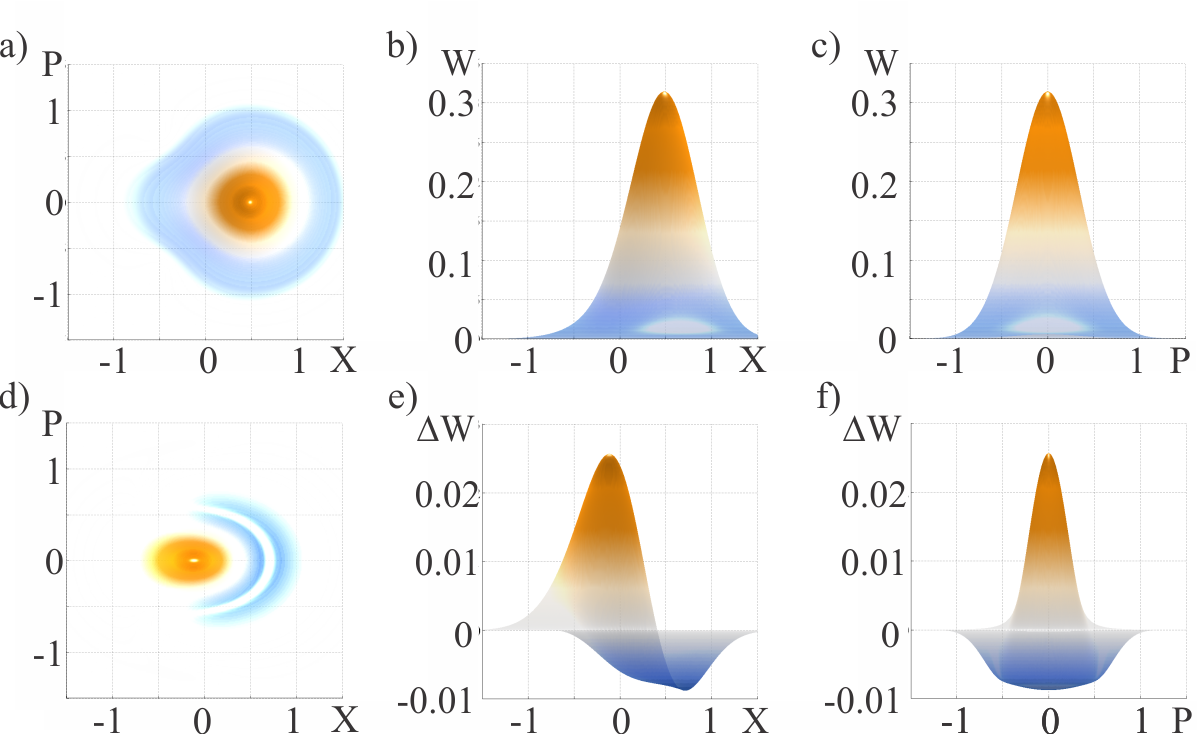}%
\caption{a)-c) Wigner function of the optimally cloned BCS for a signal with mean photon number $|\alpha|^2 = 0.5$, i.e. $\theta=0.1884$. 
The distribution is viewed from the top, as well as along the \textit{x}- and \textit{p}-quadrature.
d)-f) Illustration of the difference $W_{\mathrm{clone}}-W_{\mathrm{coherent}}$ between the optimally cloned state and the coherent signal viewed from the same perspectives.}
\label{WignerOptClone}
\end{figure}

The Wigner function of the optimally cloned state as well as the difference between the Wigner function of the initial state and the optimally cloned state are shown from different perspectives in Fig.\ref{WignerOptClone} for a signal state with mean photon number $|\alpha|^2=0.5$, i.e. $\theta=0.1884$. At these parameters the fidelity of the optimal qubit cloner is close to the minimum such that large deviations can be expected. 

The Wigner function of the optimally cloned state (upper row) exhibits a slight bias towards the opposed signal state $\ket{-\alpha}$, which is primarily visible in the top view.
This bias is clearly enhanced in the plot illustrating the difference between the Wigner function of the initial state and the optimally cloned state (lower row) . 

\begin{figure}%
\includegraphics[width=.9\columnwidth]{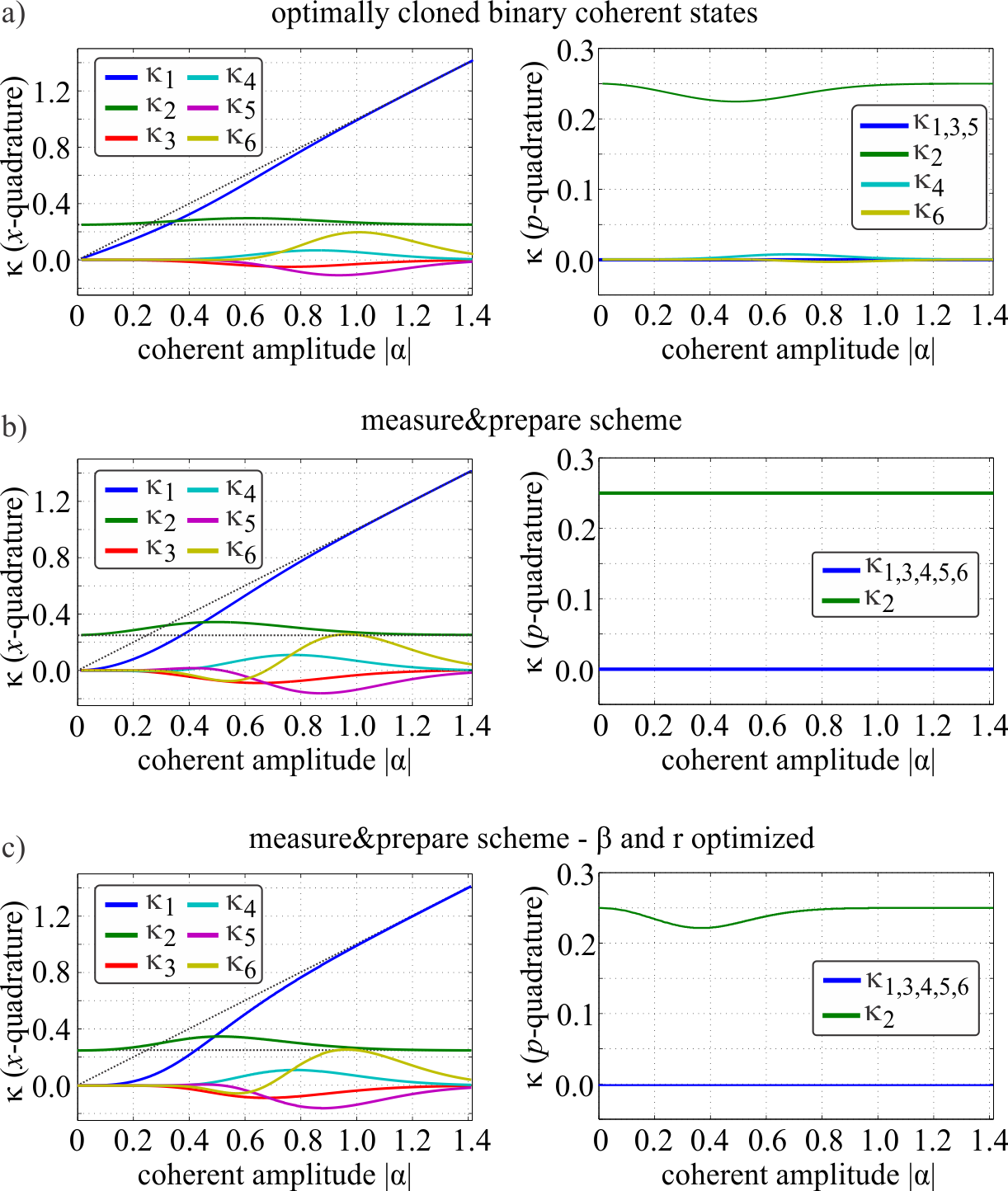}%
\caption{
a) Cumulants along the \textit{x}-quadrature and along the \textit{p}-quadrature of the optimally cloned BCS. 
The dashed black lines indicate the coherent input state's mean amplitude $\kappa_1$ and variance $\kappa_2$.
The clones are slightly squeezed along the \textit{p}-quadrature. 
All odd-numbered cumulants vanish. 
The non-vanishing kurtosis (4th cumulant) shows, that the clones exhibit non-Gaussian characteristics also along the \textit{p}-quadrature. 
This is an interesting finding that points towards the complexity of the practical realization of the optimal cloning scheme. \newline
b) Cumulants for the measure\&prepare scheme featuring the preparation of states with the exact signal amplitude $|\alpha|$, see Sec.\ref{MPC}. 
Along the \textit{x}-quadrature the variation of the cumulants is more pronounced but the overall characteristics are very similar. 
Along the \textit{p}-quadrature the only non-zero cumulant is the variance, which remains at the shot noise limit.
c) Cumulants for the measure\&prepare scheme featuring the preparation of states with optimized amplitude $|\beta|$ and optimized squeezing parameter $r$.}
\label{Cumulants}
\end{figure}

In Fig.\ref{Cumulants}a), we illustrate the first six cumulants $\kappa_n$ of the statistical distribution along the \textit{x}- and \textit{p}-quadrature for the optimally cloned BCS $\ket{+ \alpha}$.
In terms of the central moments $\mu_n = \left\langle \left( \hat{X} - \langle \hat{X} \rangle \right)^n \right\rangle$ and the mean value $m_1$ the cumulants can be expressed in a compact form as
\begin{eqnarray}
\kappa_1 &=& m_1,   \hspace{0.5cm} \kappa_2 = \mu_2 \nonumber \\
\kappa_3 &=& \mu_3, \hspace{0.5cm} \kappa_4 = \mu_4 - 3 \mu_2^2 \nonumber \\
\kappa_5 &=& \mu_5 - 10 \mu_3\mu_2 \nonumber \\
\kappa_6 &=& \mu_6 - 15 \mu_4 \mu_2 -10 \mu_3^2 + 30 \mu_2^3 \nonumber.
\label{eq:Cumulants}
\end{eqnarray} 
Gaussian distributions, like the Wigner functions of the coherent input signals, are fully determined by the first and second cumulant, i.e. the mean value and the variance. 
Hence, particularly the higher order cumulants characterize the peculiarities of the clones. 
The third and fourth cumulant are associated with the skewness and the 'tailedness' (kurtosis) of the distributions. 
Let us first discuss the cumulants along the \textit{x}-quadrature. 
The curve of the mean value $\kappa_1$ is slightly convex, i.e the mean field amplitude of the clones approaches the coherent amplitude of the input signal from below.
The variance is slightly above the Heisenberg minimum uncertainty, which in our convention is set $\kappa_2=\frac{1}{4}$. 
Moreover, we observe a negative skewness $\kappa_3<0$, i.e. the distribution is leaning to the right. 
This can intuitively be understood by considering that the cloner needs to \textit{decide} between the BCS. 
Imperfections in the cloning procedure of $\ket{+\alpha}$ results in an erroneous preparation in favor of the state $\ket{-\alpha}$ hence shifting the barycenter of the distribution towards the left while the peak remains around $\ket{+\alpha}$.

Along the \textit{p}-quadrature only even-valued cumulants contribute to the distribution. 
The complete information about the signal state is encoded along the \textit{x}-quadrature such that one could naively expect that a cloner that only interacts with the input state by partial measurements, coherent displacements and squeezing would essentially disturb the marginal distribution along this quadrature while preserving the Gaussian statistics along the \textit{p}-quadrature. 
This is exemplified in Fig.\ref{Cumulants}b)-c) where the cumulants for a measure\&prepare cloner outputting either an exact replica of one of the input states, or an optimally squeezed and optimally displaced state, are shown. These schemes are discussed in more detail in Sec.\ref{MPC}.
The non-zero contributions of the fourth and sixth cumulants $\kappa_4$ and $\kappa_6$ for the quantum-optimally cloned BCS indicate that a non-linearity of order higher than two, i.e. higher than the squeezing operation, is required to implement the optimal scheme.

\begin{figure}[b]%
\includegraphics[width=1\columnwidth]{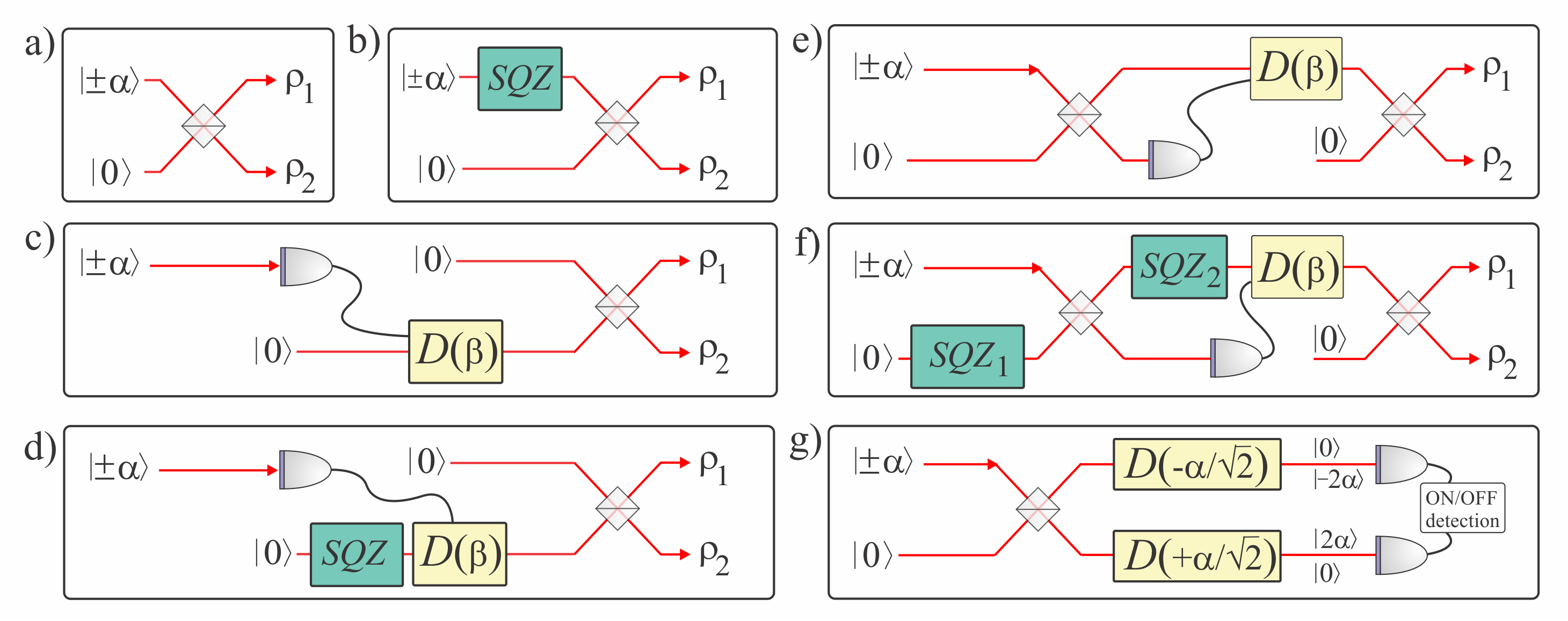}%
\caption{Overview on the different considered cloning schemes. 
a) mere beam splitting.
b) phase sensitive amplification.
c) complete measurement and preparation.
d) complete measurement, preparation and phase-sensitive amplification.
e) partial measurement and feed-forward for regeneration.
f) squeezing assisted partial measurement and feed-forward for displacement of the phase-sensitively amplified remaining state.
g) unambiguous state discrimination scheme.}
\label{CloningSchemes}%
\end{figure}

In the following we investigate different practical cloning schemes and compare their performance to the previously discussed optimal binary state cloner.
A schematic overview on the different approaches is depicted in Fig.\ref{CloningSchemes}.

\subsection{Cloning with a Single Beam Splitter}
\begin{figure}[htb]%
\includegraphics[width=.9\columnwidth]{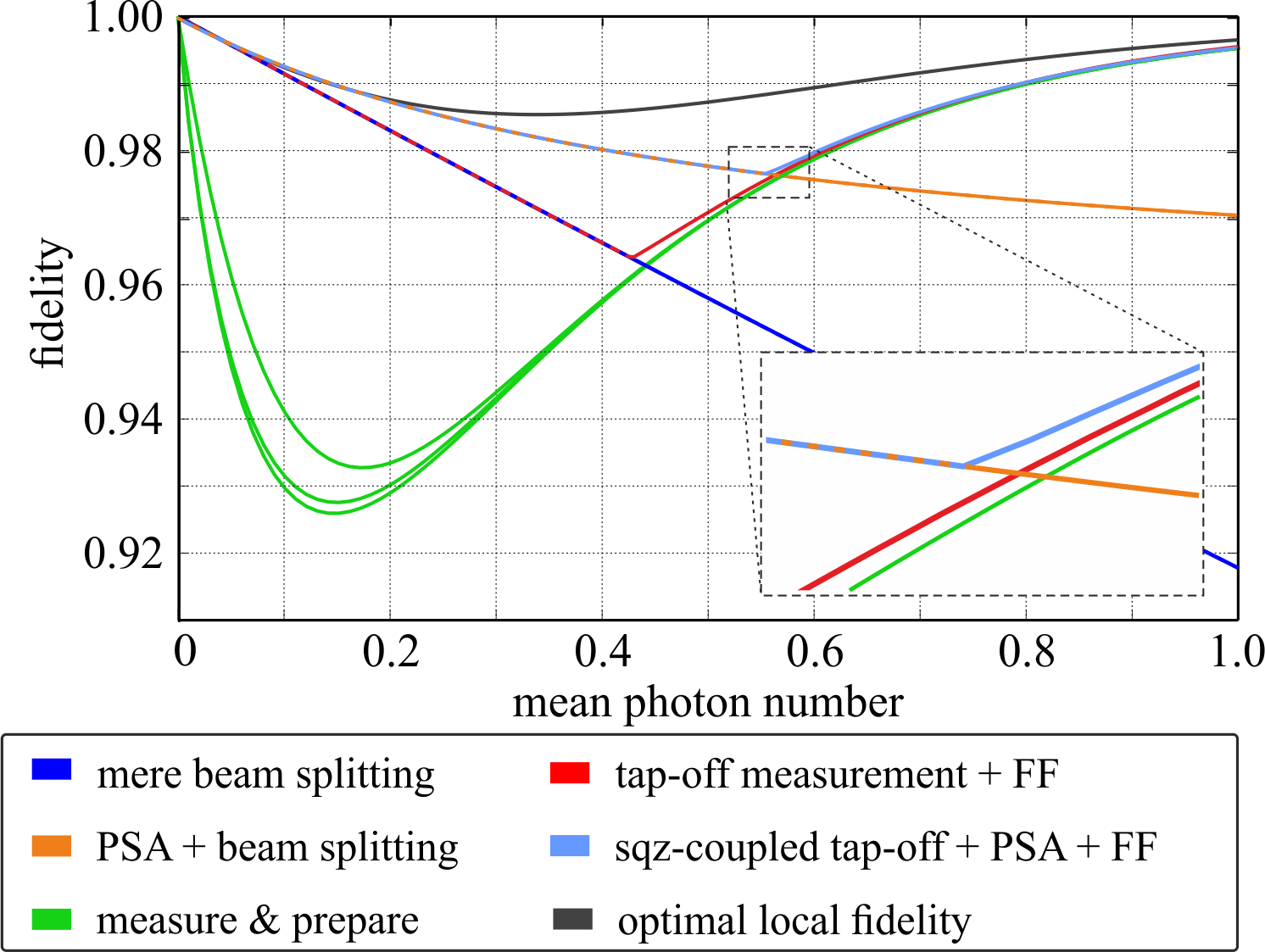}%
\caption{Comparison of the cloning fidelity for different practical cloning approaches.
See the main text for details.}%
\label{CloningFidelity}%
\end{figure}

The simplest approach to generate two copies of an unknown coherent state is to split the signal on a symmetric beam splitter, see Fig.\ref{CloningSchemes}a). 
The coherent amplitude reduces as $\alpha \mapsto \alpha/\sqrt{2}$, but the phase and the (quantum) noise characteristics are perfectly preserved. 
While obviously inadequate for bright signals, this approach is astoundingly effective in the domain of weak signals $\alpha\leq1$. 
The difference in the states' amplitude prior to and after the beam splitter is $\Delta \alpha = (1-1/\sqrt{2})\alpha$, i.e. proportional to $\alpha$. 
For small amplitudes, the absolute difference in the amplitudes is small and hence the cloning fidelity is high. 
The maximal fidelity for the pure beam splitting approach is shown as the dark blue curve in Fig.\ref{CloningFidelity}.

\subsection{Cloning via Phase-Sensitive Amplification}
Another intuitive approach to generate copies of an unknown input state is to amplify the signal to twice its initial intensity followed by symmetric beam splitting.
Conventional optical amplifiers are optimized for an unbiased processing of arbitrary signals, which requires operation in a linear and phase-insensitive way.
The conservation of the canonical commutator bracket imposes an inevitable noise penalty to the output of such amplifiers \cite{Caves1982, Ber1961, Hausxr1962}.
In the high gain regime this results in a minimal reduction of the signal-to-noise ratio by a factor of two. 
For BCS, a remedy is offered by parametric processes, in particular by the squeezing operation that is described by the squeezing operator
\begin{equation}
\hat{S}(z) = \exp\left(\frac{z^{*}}{2}\hat{a}^2 - \frac{z}{2}\hat{a}^{\dagger 2}\right),
\label{eq:SqueezingOperator}
\end{equation}
where $z = r\exp(i \phi)$ denotes the complex squeezing parameter. 
The parameter $r$ controls the degree of squeezing and the phase $\phi$ defines the squeezed field quadrature in phase space. 
A sketch of the cloner based on squeezing and subsequent beam splitting is shown in Fig.\ref{CloningSchemes}b).
The squeezing operation is a phase-sensitive amplification (PSA) and effects a hyperbolic transformation of the quadrature variables such that the mean amplitude and the quadrature variances of the BCS transform as
\begin{eqnarray}
\alpha_r &=& \cosh(r)\,\alpha, \nonumber \\
\sigma_{X,r}^{2} &=& e^{+2r}\,\sigma_{0}^{2},  \sigma_{P,r}^{2} = e^{-2r}\,\sigma_{0}^{2}. 
\label{squeezing_ampl}
\end{eqnarray}
The amplified quadrature can be aligned perfectly with the coherent states modulation quadrature (cp. Fig.\ref{FullBCS-PhaseSpace}d)) such that clones with the exact input amplitude could straightforwardly be generated. 
The resulting fidelity, however, is limited by the squeezing induced redistribution of the Heisenberg uncertainties. 
Optimization of the output fidelity thus requires to balance the trade-off between the amplification of the amplitude and the associated deviations in the quadrature variances. 
As a consequence, the optimally amplified amplitude is slightly smaller than that of the input state.
The maximal fidelity for this approach is shown as the orange curve in Fig.\ref{CloningFidelity}. 
For signal mean photon numbers up to $|\alpha|^2 \leq 0.2$ the fidelity provided by the PSA based cloner asymptotically coincides with the quantum optimal fidelity.

\subsection{Cloner with Complete Measurement and Preparation} \label{MPC}
Within the framework of classical physics an arbitrary number of perfect copies of an unknown state can be generated by precise measurement and subsequent preparation. 
In the realm of quantum mechanics, this approach is baffled by the complementarity of different observables and the potential non-orthogonality of the states in question. 
For BCS the maximal fidelity of the measure\&prepare approach is first and foremost determined by the minimal error probability in discriminating the states, i.e. the Helstrom bound, see Eq.(\ref{eq:HelstromBound}).

\begin{figure}%
\includegraphics[width=.9\columnwidth]{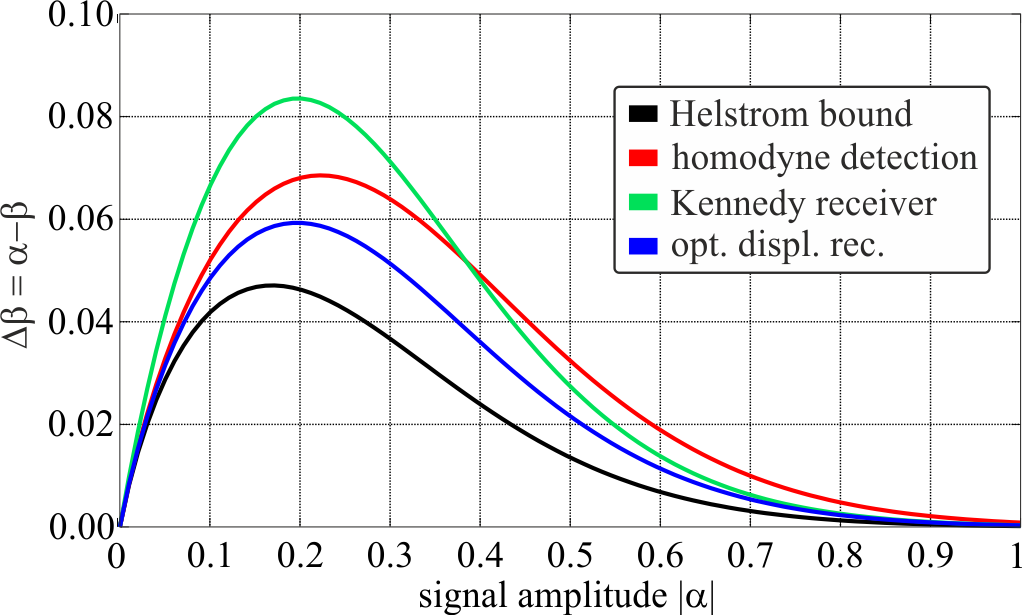}
\caption{Difference $\Delta\beta$ between the fidelity-optimized amplitude of the prepared clones $\beta$ and the signal's coherent amplitude $\alpha$ in the measure\&prepare scheme. 
Due to the finite error probability in the discrimination of the states, the optimal value is always below the amplitude of the input state and the parameters only coincide in the classical limit $\alpha\gg1$.}%
\label{MAPClonerOptAmpl}%
\end{figure}

The hypothesis acquired via the measurement determines the prepared state $\rho_{\pm}$ and the density operator for each of the clones reads
\begin{equation}
\rho^{\mathrm{clone}}_{\pm\alpha} = \left(1-p_{err}(\alpha)\right )\rho_{\pm\alpha} + p_{err}(\alpha)\rho_{\mp\alpha}. 
\label{eq:MeasureAndPrepareFidelity}
\end{equation}
For the most elementary realization of the measure\&prepare cloner the prepared state is identical to one of the coherent input states $\rho_{\pm\alpha}=\ket{\pm\alpha}\bra{\pm\alpha}$. 
The fidelity can be increased by optimizing the amplitude of the prepared coherent state $\beta$ depending on the signal amplitude $|\alpha|$ and the error probability $p_{err}(\alpha)$ of the detector.
Such a cloner is sketched in Fig.\ref{CloningSchemes}c) and the optimal amplitude values for different receiver architectures are shown in Fig.\ref{MAPClonerOptAmpl}.
The displacement parameter $\beta$  is optimized by deriving the stationary point of the fidelity.

\begin{eqnarray}
&\partial_{\beta}& \bra{\alpha} \rho_{\beta} \ket{\alpha} = 0 \nonumber \\
&\Rightarrow& \frac{\alpha + \beta}{\alpha + \beta} \exp(-4\,\alpha\,\beta)=\frac{1-p_{err}(\alpha)}{p_{err}(\alpha)}
\label{eq:MPoptimalbeta}
\end{eqnarray}
The fidelity can be further increased by preparing coherently displaced squeezed vacuum states $\rho_=\ket{\beta, r}\bra{\beta, r}$, where $\ket{\beta, r} = \hat{D}(\beta)\hat{S}(r)\ket{0}$ and optimizing both the squeezing parameter $r$ and the displacement amplitude $\beta$, see Fig.\ref{CloningSchemes}d).
The optimized fidelities for these three cases, i.e. preparation of either of the input states, preparation of an optimally displaced coherent state, and preparation of an optimally displaced and optimally squeezed state, are presented in ascending order as green curves in Fig.\ref{CloningFidelity}.

\subsection{Cloner with Partial Measurement and Preparation} \label{PPC}
Besides the complete measurement approach, it is worthwhile to consider strategies based on partial measurement and feed-forward. 
The signal is split on a beam splitter with (optimized) transmissivity $T$ and the reflected part of the state is measured with an appropriate receiver. 
The obtained information is forwarded to optimally transform the remainder of the state prior to a symmetric beam splitter generating the two clones. 
It was previously shown that such a scheme involving only feed-forward and coherent displacements saturates the fidelity bound for unconditional cloning of coherent states, $F=2/3$ \cite{Andersen2005}.
Sketches of different realizations of the partial measurement scheme are shown in Fig.\ref{CloningSchemes}e)-f).

\begin{figure}%
\includegraphics[width=1\columnwidth]{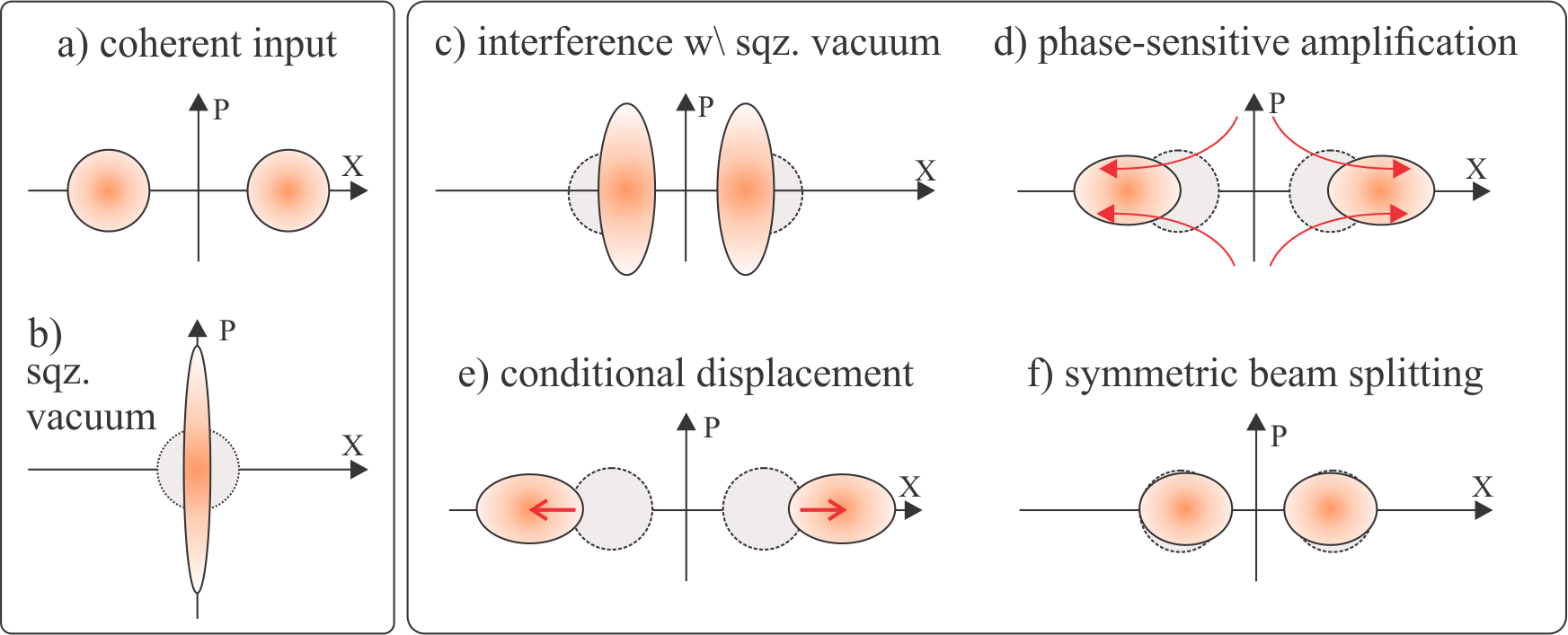}%
\caption{Phase space illustration of the BCS cloning scheme with a squeezed vacuum input, phase-sensitive amplification and a conditional displacement. 
a) BCS alphabet at the input of the cloner. 
b) Squeezed vacuum entering the open port of the tap-off beam splitter. 
c) Signal after interference with the squeezed vacuum (qualitatively equal for both the transmitted and the reflected part). 
The signal amplitude is reduced, but the states are squeezed along the $x$-quadrature, which allows for a smaller error probability in the discrimination of the reflected signals compared to a plain vacuum input. 
d) The coherent amplitude is increased via phase-sensitive amplification. 
Thereby the squeezing is shifted from the $x$-quadrature to the $p$-quadrature.
e) Subsequently, the amplitudes are further increased via coherent displacement. 
The displacement phase is conditioned on the outcome of the partial measurement on the tapped-off signal. 
f) Finally, the signal is split symmetrically on a beam splitter to generate the clones. }
\label{FullBCS-PhaseSpace}%
\end{figure}

\begin{figure}%
\includegraphics[width=.9\columnwidth]{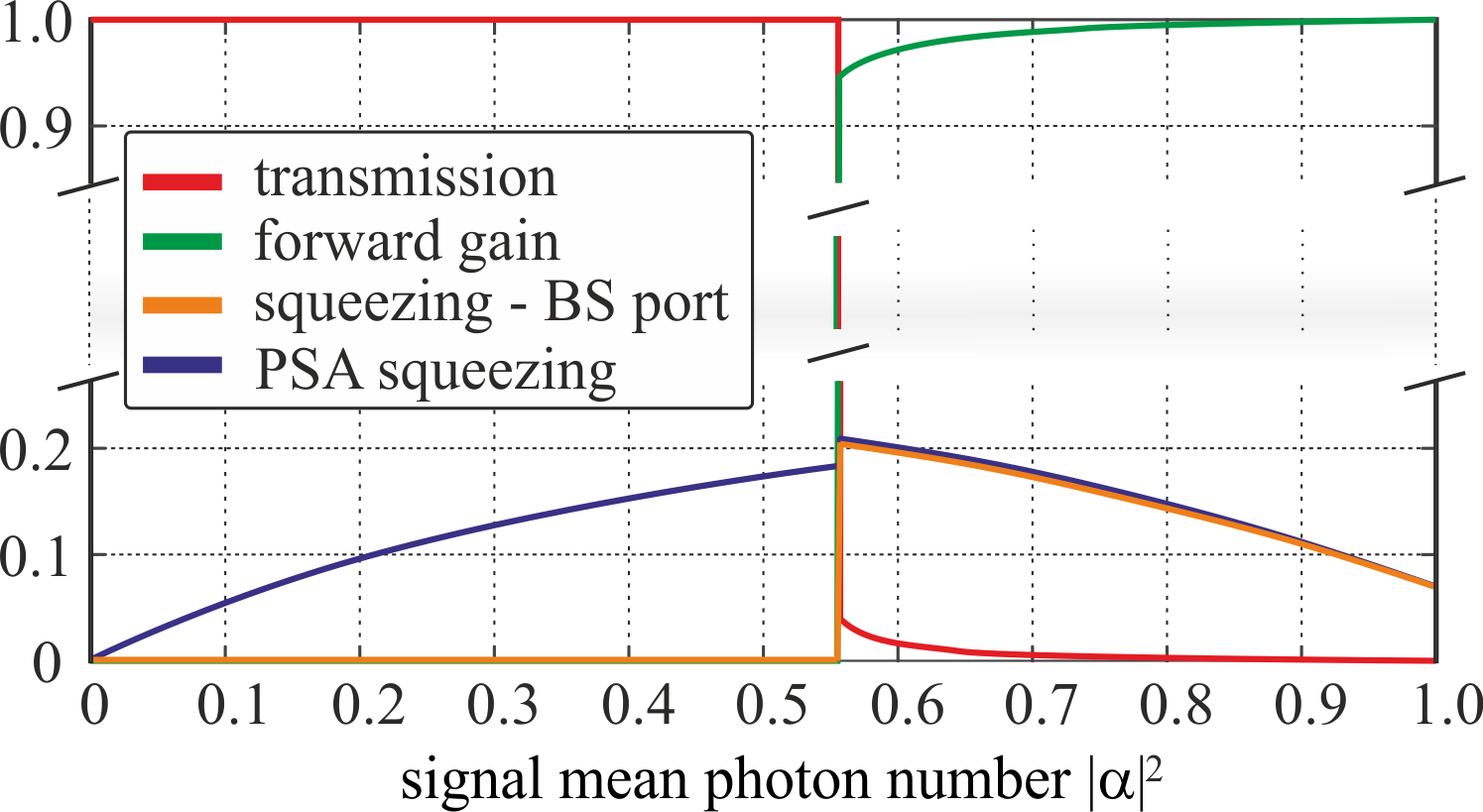}%
\caption{Optimized parameters for the cloning  scheme of Fig.\ref{CloningSchemes}f).
The corresponding y-axis labels are: transmission $T$ (red curve), forward gain $g$ (green curve), squeezing parameter $r$ (orange and purple curve). Up to a critical signal power of $|\alpha|^2 \approx 0.56$ the beam splitter is perfectly transmissive and the amplification of the coherent states is solely based on the squeezing in the phase-sensitive amplifier ($SQZ_2$, purple curve). 
For higher signal powers, the measurement provides enough information about the state to compensate for the lost signal in the tap-off. 
The transmissivity instantly drops to a value as low as $T\approx 0.05$ and asymptotically decreases to T=0. Simultaneously the forward gain to the displacement stage jumps to $g\approx 0.95$ and increases to unity. 
The PSA squeezing parameter is just slightly above the squeezing parameter for the input to the free port of the tap-off beam splitter. Both tend to zero with increasing signal power. }%
\label{full_BCS_cloner_opt_parameters}%
\end{figure}

In the splitting of the signal prior to the partial measurement, the second input port of the beam splitter was implicitly assumed to be in the vacuum state. 
However, the error probability in the discrimination of BCS can be reduced by squeezing the vacuum input in the \textit{x}-quadrature. 
Consequently, also the transmitted part of the state exhibits squeezing along the \textit{x}-quadrature hence implying a lower fidelity for this approach. 
Here, the phase-sensitive amplifier PSA discussed in the previous section comes to the rescue. 
The signal benefits from the amplification with the PSA in two ways as illustrated in Fig.\ref{FullBCS-PhaseSpace}: 
First, the state is phase-sensitively amplified in the correct direction of phase-space.
Second, the hyperbolic phase space transformation invoked by the PSA reshapes the Heisenberg uncertainty to withdraw the remaining squeezing from the squeezed vacuum input. 
Note, however, that it is not possible to regenerate a pure coherent state, as the transmitted part of the signal is already in a mixed state due to the beam splitting of the squeezed vacuum. 
By optimizing over the transmissivity of the beam splitter, the squeezing parameters, and the coherent displacement we find the fidelities presented as the light blue curve in Fig.\ref{CloningFidelity}.
The optimized parameters for the transmission $T$, the vacuum squeezing $SQZ_1$, the displacement forward gain $g$, and the squeezing in the phase-sensitive amplifier $SQZ_2$ are shown as a function of the signal's mean photon number in Fig.\ref{full_BCS_cloner_opt_parameters}.
The forward gain is defined via 
\begin{equation}
\beta = g\, \left( \sqrt{2} - T \right)\, \alpha. 
\label{eq:forward_gain}
\end{equation}
The factor $\sqrt{2}$ accounts for the beam splitting at the output of the cloner. 
Unit gain corresponds to the preparation of signal states with the exact amplitude 
of the input states. \\
Up to a mean photon number of $|\alpha|^2 \approx 0.55$ the fidelity is maximized by mere phase-sensitive amplification, i.e. the tap off beam splitter is completely transmissive. 
Above this threshold the transmissivity drops almost to zero and the forward gain jumps close to unity. 
The optimized squeezing parameter for the squeezed vacuum input and for the PSA almost coincide but the optimized PSA squeezing is slightly higher. 
By further increasing the signal power the parameters approach the classical scenario; i.e. no squeezing, complete measurement of the state and preparation of the clones with unit gain. 

\subsection{Cloning via Unambiguous State Discrimination}\label{USDCloning}
Unambiguous state discrimination USD (also known as Ivanovic-Dieks-Peres measurement \cite{Ivanovic1987, Dieks1988, Peres1988}) is a generalized measurement allowing for the perfect identification of an unknown quantum state from a known alphabet.
Owing to the potential non-orthogonality of the alphabet, perfect identification generally comes at the expense of a finite success probability, which for the binary coherent alphabet is upper bounded by $p_{succ} \leq 1-|\bra{-\alpha}\,\alpha\rangle|= 1-\exp(-2|\alpha|^2)$. 

A simple experimental scheme achieving this bound is sketched in Fig.\ref{CloningSchemes}g). 
The signal $\ket{\pm \alpha}$ is divided on a symmetric beam splitter and the emerging fields $\ket{\pm \alpha/\sqrt{2}}$ are displaced such that the signal with positive/negative sign is displaced to the vacuum state $\ket{0}$ and $\ket{+\sqrt{2}\alpha}$, respectively. 
As the vacuum state is an eigenstate of the photon number basis with eigenvalue zero, the detection of one or more photons in either of the detectors unambiguously identifies the input state. 
The photon detection probability is $p_{click} = 1- p_{vac} = 1- \exp(-2|\alpha|^2)$, which satisfies the USD bound. 

\begin{figure}%
\includegraphics[width=1\columnwidth]{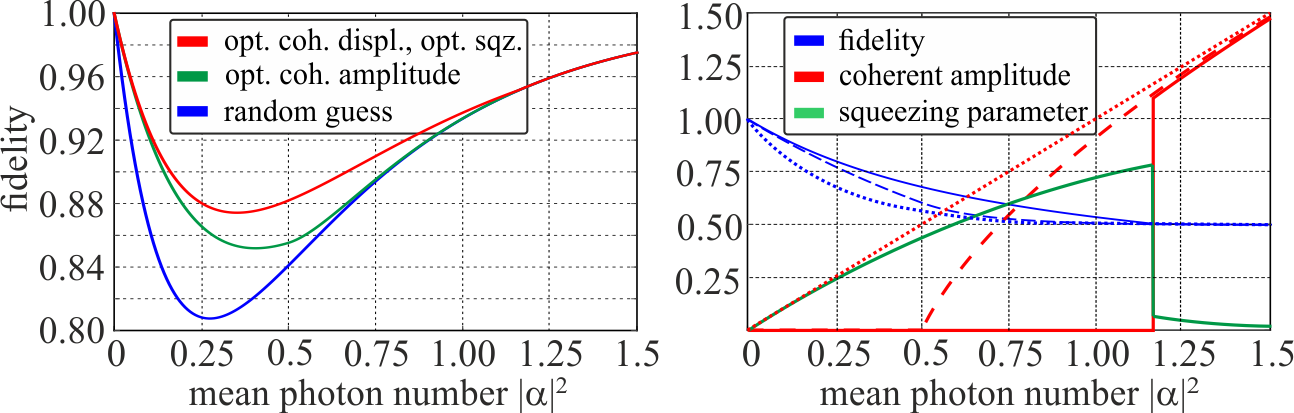}%
\caption{Output fidelity and optimized parameters for the BCS cloning with an unambiguous state discrimination scheme.  
left) Output fidelities for differently prepared signal states in case of an inconclusive outcome. 
(blue) randomly preparing either of the input states, 
(green) preparing a coherent state with optimized amplitude and 
(red) preparing a squeezed coherent state with optimized amplitude and squeezing parameter. 
right) the optimized parameters associated with the curves on the left and the separate fidelities for the inconclusive outcomes. 
(dotted) randomly preparing either of the input states.
(dashed) optimized coherent state without squeezing.
(solid) optimized squeezed coherent state.}%
\label{USD_opt_beta}%
\end{figure}

Perfectly identified signal states can straightforwardly be regenerated and contribute with unit fidelity $\mathrm{F}=1$. 
Inconclusive outcomes that occur with probability  $p_{inc} = 1-p_{succ}$ do not provide any information about the input state.
A naive approach is to prepare either of the potential signals at random, providing perfect fidelity in $50\%$ of the cases and a residual fidelity of $\mathrm{F} = \exp(-4|\alpha|^2)$ whenever the false state was prepared. 
Similar to the previously discussed situation for the full/partial measure\&prepare cloners, one can prepare a fidelity optimized state that does, in general, not coincide with any of the signal states. 
For the latter case, we again consider both the preparation of an optimized coherent state and the preparation of an optimized displaced squeezed vacuum state.
The results are combined in Fig.\ref{USD_opt_beta}. 
For the fidelity-optimized coherent states the vacuum state maximizes the fidelity up to a mean photon number of $|\alpha|^2=0.5$. 
For brighter signals the optimized amplitude asymptotically approaches the original signal amplitude. 
The resulting fidelity can be further increased by taking squeezed states into account. 
In this case, squeezed vacuum (where the squeezing is in the quadrature orthogonal to the signal states' coherent excitation) maximizes the fidelity up to a critical mean photon number of about $|\alpha|^2\approx 1.33$. 
At this power, the curve is unsteady and for higher amplitudes a state with coherent amplitude but significantly less squeezing maximizes the fidelity. 
The optimized fidelity and squeezing parameters are shown in Fig.\ref{USD_opt_beta}.
In comparison to the other discussed cloning strategies, however, the performance of the USD scheme ranks behind.

\section{Conclusion}
The formal analogy between the Hilbert spaces of qubits and of the binary coherent state alphabets allows for an analysis of the cloning fidelity of the binary coherent states within the Bloch sphere formalism. 
In this article, we put this peculiarity into practice to investigate the characteristics of the optimally cloned BCS states.
Building up on a scheme for the optimal unconditional cloning of coherent states, we proposed and analysed different
practical cloning schemes and compared their performance to the optimal cloner. 
It remains an open task to device an experimental setup that yields binary coherent state clones satisfying the quantum-optimal fidelity.
The analysis of practical cloning schemes for the four-partite quadrature phase-shift keying alphabet (QPSK) as well as minimum disturbance measurements and optimized teleportation schemes for the binary coherent state alphabet are subject of ongoing research.

\section*{Acknowledgments} 
We acknowledge financial support from the Danish Council for Independent Research (Sapere Aude grant no. 10-081599).

\bibliographystyle{unsrt}

\end{document}